\documentclass[11pt]{article}
\usepackage{hyperref}
\usepackage{color}
\usepackage{amsmath}
\pdfoutput=1
\begin{document}
\title{Bi-disperse particle-laden flows in the Stokes regime}

\author{G. Urdaneta, K. Allison, T. Crawford, S. Meguerdijian\\ W. Rosenthal, S. Lee, A. Mavromoustaki, A. L. Bertozzi  \\
\\\vspace{6pt} Department of Mathematics \\ University of California Los Angeles}
\maketitle
\begin{abstract}
This arXiv article describes the fluid dynamics video on `Bi-disperse particle-laden flows in the Stokes regime', presented at the 65th Annual Meeting of the APS Division of Fluid Dynamics in San Diego, CA in November 2012. The video shows three different experiments which aim to investigate the dynamics of a thin film of silicone oil, laden with glass beads and the effects of adding a second species of particles to the slurry. The mixture of oil and particles is allowed to flow down an incline under the action of gravity. The videos were recorded at the UCLA Applied Math Laboratory.

\end{abstract}
\section{Video technical details}
The video (Entry \# 83832, APS 65$^{\text{th}}$ Annual DFD Meeting 2012) demonstrates the patterns exhibited by gravity-driven, particle-laden thin films flowing down
a solid substrate. The results from three experiments are shown in the video. A finite volume of fluid mixed
with particles is allowed to flow down the solid plane;
visualization is achieved from the front view and the videos have
been recorded with a digital SLR camera (Canon EOS Rebel T2i). The suspension consists of silicone oil, glass and ceramic beads. The ceramic beads are denser than the glass beads while both sets of particles are denser than the fluid. It is noted that both species of beads are of the same size. The substrate angle of inclination is fixed at $\alpha=30^{o}$ while the total particle concentration, $\phi_{\text{total}}$ is fixed at $\phi_{\text{total}}=0.40$. The series of experiments conducted aim in understanding the effect of adding a second, heavier species of beads to a slurry composed of oil and glass beads. In order to visualize the separation, if any, between the two species, the ceramic and glass beads are dyed blue and red, respectively. We introduce a dimensionless parameter, $\eta$, defined as the ratio of the concentration of glass beads to the total concentration of particles.

The rightmost video shows a reference case wherein the suspension consists only of glass beads (i.e. $\eta =1$); the relative ratio $\eta$ is decreased from 1 (right) to 0.75 (center) to 0.25 (left) with the addition of ceramic beads. The runtime associated with
the individual videos has been fast-forwarded 6 times, in order to
demonstrate the onset of flow instabilities as well as the development of fingering patterns. The choice of parameters allows the presentation of three, distinct
regimes, also exhibited by an increase in particle concentration in monodisperse slurry flows. In this video, we observe the three regimes by keeping the total particle concentration constant while we add a second species of negatively buoyant beads. For large concentrations of \textit{ceramic} beads (left video, mostly blue
particles), the particles settle rapidly allowing the clear fluid to
flow over them which results in fingering. For small concentrations of \textit{ceramic} beads (middle video, red and blue particles), we observe a well-mixed regime characterized
by finger formation; this regime is considered to give an unstable, transient pattern. Finally, for monodisperse suspensions of \textit{glass} beads i.e. no \textit{ceramic} beads (right video, red particles), the beads aggregate at the
contact line, forming a particle-rich ridge, evident by a darker red color at the front of the flow. \\ \ \\ 

\noindent \textbf{References}\ \\ \

\noindent 1. T. Ward, C. Wey, R. Glidden, A. E. Hosoi, and A. L. Bertozzi. Experimental study of gravitation effects in the flow of a particle-laden thin film on an inclined plane, \textit{Phys. Fluids}, \textbf{21}, 083305 (2009).\\

\noindent 2. B. Cook,  O. Alexandrov and A. L. Bertozzi. Linear stability of particle-laden thin films, \textit{The European Physical Journal - Special Topics}, \textbf{166}, 1, 77-81 (2009).\\

\noindent 3. N. Murisic , J. Hob, V. Huc, P. Latterman, T. Koche, K. Linf, M. Mata, A.L. Bertozzi. Particle-laden viscous thin-film flows on an incline: Experiments compared with a theory based on shear-induced migration and particle settling, \textit{Physica D: Nonlinear Phenomena}
\textbf{240}, 20, 1661-1673 (2011). \\

\noindent 4.N. Murisic, B. Pausader, D. Peschka, A.L. Bertozzi. Dynamics of particle settling and resuspension in viscous
liquids,
\textit{under review for J. Fluid Mech}. \\ \ \\ 

\noindent \textbf{Acknowledgements}\ \\ \

\noindent The authors would like to thank Miss Kaiwen Huang for her help in conducting the experiments presented in this video.

\end{document}